# Analysis of quantum light memory in atomic systems

Hassan Kaatuzian [1]; Ali Rostami [2]; Ahmad Ajdarzadeh Oskouei [1],*

1:Dept. of Electrical Eng. AmirKabir University of Technology, Tehran, Iran
2:Faculty of Electrical Eng. , University of Tabriz, Tabriz, 51664, Iran
E-Mail: hsnkato@cic.aut.ac.ir, rostami@tabrizu.ac.ir, ahmad_ajdarzadeh@engineer.com

We extend the theory of quantum light memory in atomic ensemble of $\Lambda$ type atoms with considering $\gamma_{bc}$ (lower levels coherence decay rate) and one and two-photon detunings from resonances in low intensity and adiabatic passage limit. We obtain that with considering these parameters, that there will be a considerable decay of probe pulse and stored information; also, we obtain that the group velocity of probe (light) pulse and its amplitude does not tend to zero by turning off the control field. We propose a method to keep the probe pulse in small values in turn off time of control field and to reduce the loss of the stored probe pulse. In addition, we obtain that in the Off-resonance case there will be a considerable distortion of the output light pulse that causes in loss of the stored information, then we present limitations for detunings and therefore for bandwidths of practical lasers also limitations for maximum storage time to have negligible distortion of stored information. We finally present the numerical calculations and compare them with analytical results.

## 1. Introduction

Atomic coherence and related phenomena such as Electromagnetically Induced Transparency (EIT) and Slow Light have been studied extensively in recent years [1-10]. Many application are proposed to this phenomena such as nonlinear optics (SBS, FWM and etc.), Lasing without Inversion, Laser cooling and Sagnac Interferometer [11-14]. One of important and promising applications in this field is light storage and quantum light memory that is investigated by some research groups [15-26]. The most common mechanism in this application is that the light pulse is trapped and stored in atomic excitations in the EIT medium by turning off the control field and then it is released by turning on the control field. Most of these works do not present a clear and general theory to analyze the propagation and storage of light. In addition, most of the works in EIT and slow light treat the light classically that is not proper to extend to quantum memory in which quantum state of light is to be stored. The most general theory for quantum memory was developed by M. Fleischhauer, et. al. [16,17]. They consider the light, quantum mechanically and present an excellent theory to describe the case. However, their work is not general, from our point of view, in some cases. The most deficient aspect of their work is that they do not consider the decay rate of lower levels coherence and the detuning from resonances, which have important effects on the propagation and storage of light in the atomic media. It has caused their theory to be ideal and inexact. In this paper, we try to extend the theory, previously developed in Ref [16,17] to a more general and clear quantum mechanical theory for slow light and light storage in atomic ensemble with considering all decay rates and detunings and as the result, we reach to important properties about this type of quantum memory.

The organization of this paper is as follows. In section II, quantum mechanical model to describe slow light and light storage is presented. In this section after introducing the mathematical model, two subsections including low intensity limit and slow variations and adiabatic passage limit are discussed and a proposition to turning off the control field is given. The results and discussion is presented in section III. Also, in this section we have two subsections as resonance and off resonance conditions. Numerically simulated results are given in section IV. Finally, the paper is ended with a short conclusion.

## 2. Quantum mechanical model to describe slow light and light storage

The atomic system that we consider is $\Lambda$ type three level atoms, which is demonstrated in Fig. (1). The probe field couples the two $|a\rangle$ and $|b\rangle$ atomic levels to each other and the respected detuning is defined as $\omega_{ab} - v_p = \Delta + \Delta_p$. Also the control (coupling) field couples the two $|a\rangle$ and $|c\rangle$ levels with a detuning from resonance defined as

---
*Material presented in this paper is a part of Ahmad Ajdarzadeh Oskouei's work on his thesis towards M.Sc. degree. Dr. Hassan Kaatuzian and Dr. Ali Rostami are his first and second advisors on thesis.

$\omega_{ac} - \nu_c = \Delta$, where $\nu_\mu$ are related to the probe and the control field carrier frequencies and $\omega_{\alpha\beta}$ are the resonance frequencies of corresponding levels. $\Delta$ and $\Delta_p$ are defined as one and two photon detuning respectively.

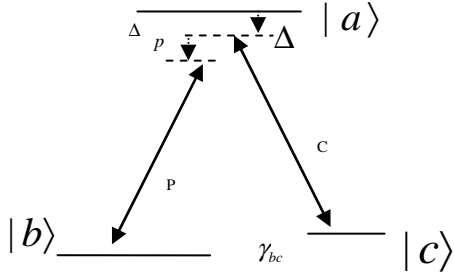

Fig. (1): Schematics of $\Lambda$ type three level atoms

The probe field $\hat{E}(z,t)$ can be defined as follows [16]

$$\hat{E}(z,t) \equiv \sqrt{\frac{\hbar \nu}{2\epsilon_0 V}} \hat{\mathcal{E}}(z,t) \times e^{i\frac{\nu_p}{c}(z-ct)} \quad (1)$$

In this relation $\hat{\mathcal{E}}(z,t)$ is the slowly varying annihilation operator (dimensionless field operator) that corresponds to the envelope of probe field. $V$ is the quantization volume of the field that can be chosen equal to the volume of the memory cell. The atomic operator for the j-th atom is defined as

$$\hat{\sigma}^j_{\alpha\beta} \equiv |\alpha_j\rangle\langle\beta_j|. \quad (2)$$

In this relation $|\alpha_j\rangle$ and $|\beta_j\rangle$ are the Heisenberg Picture base kets (States) for j-th atom. We can divide the memory cell to sections in which atomic operator does not change and every section is characterized by coordinate z. In this way, one can define the collective (continuum) atomic operators as [16,27]

$$\hat{\sigma}_{\alpha\beta}(z,t) \equiv \frac{1}{N_z}\sum_{j=1}^{N_z} \hat{\sigma}^j_{\alpha\beta}, \quad (3)$$

where $N_z$ is the number of atoms in the section z. For our purposes, it is easier to work with slowly varying collective atomic operators that are defined as

$$\hat{\tilde{\sigma}}_{\alpha\beta}(z,t) \equiv \hat{\sigma}_{\alpha\beta}(z,t)e^{i\frac{\omega_{\alpha\beta}}{c}(z-ct)} \quad (4)$$

With considering these operators the interaction Hamiltonian in interaction picture can be written as [28]

$$\hat{H}_I = -N\int \frac{dz}{L}(\hbar g \hat{\mathcal{E}}(z,t)\hat{\tilde{\sigma}}_{ab}(z,t)e^{i(\Delta+\Delta_p)t} + \hbar\Omega\hat{\tilde{\sigma}}_{ac}(z,t)e^{i\Delta t}) + H.a. \quad (5)$$

where $H.a.$ refers to Hermitian adjoint of the integral, $N$ is the total number of atoms in the memory cell, $L$ is the length of the cell, and g is the vacuum Rabi frequency given by $g = \frac{\vec{\wp}_{ij}\cdot\vec{\in}\sqrt{\frac{\hbar\nu_p}{2\epsilon_0 V}}}{\hbar}$ that is related to atom-field coupling strength in a given interaction system ($\vec{\wp}_{ij}$ is the electric dipole moment corresponding to the two levels i and j, and $\vec{\in}$ is the field polarization.). $\Omega$ is defined as the Rabi frequency of control field that is given by $\Omega = \vec{\wp}_{ac}\cdot\vec{E}_c/\hbar$, where $\vec{E}_c$ is the amplitude of the control field. One can find equations of motion for the atomic and field operators by substituting the above Hamiltonian in the Heisenberg-Langevin equations [27-30] as

$$[\frac{\partial}{\partial t} + c\frac{\partial}{\partial z}]\hat{\mathcal{E}}(z,t) = igN\hat{\tilde{\sigma}}_{ba}(z,t) \quad (6)$$

$$\frac{\partial}{\partial t}\hat{\tilde{\sigma}}_{bc}(z,t) = -(i\Delta_p + \gamma_{bc})\hat{\tilde{\sigma}}_{bc} - ig\hat{\mathcal{E}}(z,t)\hat{\tilde{\sigma}}_{ac}$$
$$+ i\Omega^*\hat{\tilde{\sigma}}_{ba} + \hat{F}_{bc}(z,t) \quad (7a)$$

$$\frac{\partial}{\partial t}\hat{\tilde{\sigma}}_{ba}(z,t) = -(i(\Delta+\Delta_p)+\gamma_{ba})\hat{\tilde{\sigma}}_{ba}$$
$$+ ig\hat{\mathcal{E}}(z,t)(\hat{\tilde{\sigma}}_{bb}-\hat{\tilde{\sigma}}_{aa})+i\Omega\hat{\tilde{\sigma}}_{bc}$$
$$+\hat{F}_{ba}(z,t) \quad (7b)$$

$$\frac{\partial}{\partial t}\hat{\tilde{\sigma}}_{ca}(z,t) = -(i\Delta+\gamma_{ca})\hat{\tilde{\sigma}}_{ca}+i\Omega(\hat{\tilde{\sigma}}_{cc}-\hat{\tilde{\sigma}}_{aa})$$
$$+ig\hat{\mathcal{E}}(z,t)\hat{\tilde{\sigma}}_{cb}+\hat{F}_{ca}(z,t) \quad (7c)$$

$$\frac{\partial}{\partial t}\hat{\tilde{\sigma}}_{aa}(z,t) = -\gamma_a\hat{\tilde{\sigma}}_{aa}-ig[\hat{\mathcal{E}}^+(z,t)\hat{\tilde{\sigma}}_{ba}-H.a.]$$
$$-i[\Omega^*\hat{\tilde{\sigma}}_{ca}-H.a.]+\hat{F}_a(z,t) \quad (7d)$$

$$\frac{\partial}{\partial t}\hat{\tilde{\sigma}}_{bb}(z,t) = \frac{\gamma_a}{2}\hat{\tilde{\sigma}}_{aa}+\gamma_c\hat{\tilde{\sigma}}_{cc}$$
$$+ig[\hat{\mathcal{E}}^+(z,t)\hat{\tilde{\sigma}}_{ba}-H.a.]+\hat{F}_b(z,t) \quad (7e)$$

$$\frac{\partial}{\partial t}\hat{\tilde{\sigma}}_{cc}(z,t) = \frac{\gamma_a}{2}\hat{\tilde{\sigma}}_{aa}-\gamma_c\hat{\tilde{\sigma}}_{cc}+i[\Omega^*\hat{\tilde{\sigma}}_{ca}-H.a.]$$
$$+\hat{F}_c(z,t) \quad (7f)$$

The sign (+) on operators is the Dagger sign that correspond to Hermitian adjoint of the operators. $\gamma_\alpha$ and $\gamma_{\alpha\beta}$ are the population decay rate of level $\alpha$ and the decay rate of coherence between levels $\alpha$ and $\beta$ respectively. Also, $\hat{F}_\alpha, \hat{F}_{\alpha\beta}$ are $\delta$ correlated Langevin noise operators that are caused by reservoir noise fluctuations (Vacuum Modes) [16,27,29]. In the above equations we have included the $\gamma_{bc}, \Delta, \Delta_p$ terms which are ignored in the main reference [16]. These parameters, as we will show, have considerable effects on the memory behavior. The present equations are a set of coupled differential equations and it is difficult to solve them. Therefore, we use some approximations to minimize these equations.

## A. Low intensity limit

For the first approximation, we assume low intensity approximation in which the probe field is very weak compared to the control field. With this approximation, one can consider $\hat{\mathcal{E}}$ as a perturbation in above equations and can reach to below relations in the first order of approximation [16,29].

$$<\hat{\tilde{\sigma}}_{bb}(z,t)> \cong 1$$
$$<\hat{\tilde{\sigma}}_{aa}(z,t)>, <\hat{\tilde{\sigma}}_{cc}(z,t)>, <\hat{\tilde{\sigma}}_{ac}(z,t)> \cong 0 \quad (8)$$
$$<\hat{\tilde{\sigma}}_{ba}(z,t)>, <\hat{\tilde{\sigma}}_{bc}(z,t)> \neq 0 \quad (small)$$

The Eq. (7) can then be reduced to the following equations as

$$\frac{\partial}{\partial t}\hat{\tilde{\sigma}}_{bc} = -(i\Delta_p + \gamma_{bc})\hat{\tilde{\sigma}}_{bc} + i\Omega^*\hat{\tilde{\sigma}}_{ba} + \hat{F}_{bc}, \quad (9a)$$

$$\frac{\partial}{\partial t}\hat{\tilde{\sigma}}_{ba} = -(i(\Delta+\Delta_p) + \gamma_{ba})\hat{\tilde{\sigma}}_{ba} + ig\hat{\mathcal{E}}$$
$$+ i\Omega\hat{\tilde{\sigma}}_{bc} + \hat{F}_{ba}. \quad (9b)$$

Eq. (9) can be rewritten in the following form as

$$\hat{\tilde{\sigma}}_{ba} = -\frac{i}{\Omega^*}[(i\Delta_p + \gamma_{bc} + \frac{\partial}{\partial t})\hat{\tilde{\sigma}}_{bc} - \hat{F}_{bc}] \quad (10a)$$

$$\hat{\tilde{\sigma}}_{bc} = -\frac{i}{\Omega}(i(\Delta+\Delta_p) + \gamma_{ba} + \frac{\partial}{\partial t})\hat{\tilde{\sigma}}_{ba}$$
$$- \frac{g}{\Omega}\hat{\mathcal{E}} + \frac{i}{\Omega}\hat{F}_{ba}. \quad (10b)$$

For simplicity and clarity of equations and solutions, field-atomic operators are converted to Dark and Bright state operator equations. The Dark and Bright state operators are defined respectively as [16]

$$\hat{\Psi}(z,t) \equiv \cos\theta \times \hat{\mathcal{E}}(z,t) - \sqrt{N}\sin\theta \times \hat{\tilde{\sigma}}_{bc}(z,t), \quad (11)$$
$$\hat{\Phi}(z,t) \equiv \sin\theta \times \hat{\mathcal{E}}(z,t) + \sqrt{N}\cos\theta \times \hat{\tilde{\sigma}}_{bc}(z,t)$$

where $\hat{\Psi}(z,t)$ is named as the Polariton operator and is a superposition of field and atomic operators. As we will see, this operator defines the propagation and storage of Information in the medium. $\theta$ is the control field strength parameter and is defined as

$$\tan\theta \equiv \frac{g\sqrt{N}}{\Omega}. \quad (12)$$

Other relations for $\theta$ can be written as

$$\cos\theta = \frac{\Omega}{\sqrt{\Omega^2 + g^2N}},$$
$$\sin\theta = \frac{g\sqrt{N}}{\sqrt{\Omega^2 + g^2N}}. \quad (13)$$

In the above equations, $\theta$ is a function of time ($\theta(t)$). When the control field is strong enough, $\theta$ tends to zero and when the control field is weak or is turned off, $\theta$ tends to $\pi/2$. It can be verified that expressions for $\hat{\mathcal{E}}(z,t)$ and $\hat{\tilde{\sigma}}_{bc}(z,t)$, with respect for the Dark and Bright state operators are as follows

$$\hat{\tilde{\sigma}}_{bc} = -\frac{1}{\sqrt{N}}(\sin\theta \times \hat{\Psi} - \cos\theta \times \hat{\Phi}), \quad (14a)$$

$$\hat{\mathcal{E}} = \cos\theta \times \hat{\Psi} + \sin\theta \times \hat{\Phi}. \quad (14b)$$

We now get back to Eq. (6) to derive a differential equation for $\hat{\Psi}$ and $\hat{\Phi}$. By substituting Eq. (10a) in Eq. (6) we obtain the following equation as

$$[\frac{\partial}{\partial t} + c\frac{\partial}{\partial z}]\hat{\mathcal{E}}(z,t) = igN(-\frac{i}{\Omega^*}[(i\Delta_p + \gamma_{bc} + \frac{\partial}{\partial t})$$
$$\hat{\tilde{\sigma}}_{bc} - \hat{F}_{bc}]). \quad (15)$$

By substituting Eq. (14) into the Eq. (15) and doing some mathematical manipulation, we obtain a differential equation in terms of $\hat{\Psi}$ and $\hat{\Phi}$ as

$$\frac{\partial}{\partial t}\hat{\Psi} + c\cos^2\theta\frac{\partial}{\partial z}\hat{\Psi} + (i\Delta_p + \gamma_{bc})\sin^2\theta \times \hat{\Psi}$$
$$= -\hat{\Phi}\dot{\theta} - c\sin\theta\cos\theta\frac{\partial}{\partial z}\hat{\Phi}$$
$$+ (i\Delta_p + \gamma_{bc})\sin\theta\cos\theta \times \hat{\Phi} - \frac{gN}{\Omega}\cos\theta \times \hat{F}_{bc}. \quad (16)$$

In deriving the above equation, we have assumed $\Omega$ to be real. In addition, the following assumption is made.

$$\frac{\partial}{\partial z}\Omega = 0 \Rightarrow \frac{\partial}{\partial z}\theta = 0 \quad (17)$$

This assumption is reasonable because in low intensity approximation, most of the population is in the ground state $|b\rangle$ and therefore, velocity of control pulse is about the speed of light in vacuum. Substituting Eq. (14) in Eq. (10b) and using Eq. (10a) yields to another equation For $\hat{\Psi}$ and $\hat{\Phi}$. After doing some manipulation we obtain the following equation as

$$\hat{\Phi} = \frac{\sin\theta}{g^2N}(i(\Delta+\Delta_p) + \gamma_{ba} + \frac{\partial}{\partial t})(\tan\theta(i\Delta_p + \gamma_{bc} + \frac{\partial}{\partial t}))$$
$$(\sin\theta \times \hat{\Psi} - \cos\theta \times \hat{\Phi}) + i\frac{\sin\theta}{g}\hat{F}_{ba}. \quad (18)$$

Eqs. (16),(18) are the two general equations to describe the propagation of $\hat{\Psi}$ and $\hat{\Phi}$ in low intensity limit.

## B. Slow variations and Adiabatic passage limit

In order to achieve more simple equations, we assume adiabatic passage limit, which means time variations are small, so that the system have enough time to set itself within Dark state. The conditions for adiabatic passage limit is discussed so far [5,16,31,32] and given as

$$L_p \gg \sqrt{\frac{\gamma_{ba}cL}{g^2N}}, \quad (19a)$$

$$T_r \gg \frac{\gamma_{ba}}{g^2N}\frac{v_{g0}}{c}, \quad (19b)$$

where $L_p$ is the length of input probe pulse in the medium and L is the total length of memory cell. $T_r$ is characteristic time corresponding to duration of turning on and off of the control field. $v_{g0}$ is the initial group velocity of probe pulse after entering the medium. Eq. (19a) corresponds to adiabatic propagation of light pulse in the medium and it means that the bandwidth of input pulse must be small compared to the transparency window of medium. Eq(19b) corresponds to adiabatic rotation of $\theta$ (turning on and off the control field) that is usually fulfilled in practical situations. In adiabatic limit, Langevin noise operators are negligible because they are $\delta$ correlated [16]. In order to apply the adiabatic passage limit to propagation equations, an adiabatic parameter is defined as

$$\epsilon \equiv \frac{1}{g\sqrt{NT}} \qquad (20)$$

where T is a characteristic time corresponding to the probe pulse duration and turn-off and turn on durations. We can imagine $\frac{1}{T} \approx \frac{\partial}{\partial t}$ and then replace $\frac{\partial}{\partial t}$ by ($g\sqrt{N}\epsilon$) in Eq. (18). In the zeroth order of $\epsilon$ ($\epsilon = 0$) that corresponds to adiabatic passage limit, we reach to a simple relation between $\hat{\Psi}$ and $\hat{\Phi}$ as below

$$\hat{\Phi} = \frac{(i(\Delta+\Delta_p)+\gamma_{ba})(i\Delta_p+\gamma_{bc})\tan\theta\sin^2\theta}{g^2N+(i(\Delta+\Delta_p)+\gamma_{ba})(i\Delta_p+\gamma_{bc})\sin^2\theta}\hat{\Psi}. \qquad (21)$$

From this equation, it can be inferred that when $\Delta_p$ or $\gamma_{bc}$ is not equal to zero, there will be a population in the bright state ($\hat{\Phi} \neq 0$) that causes a decay of input pulse (information). It is notable that in previous work [16], they have obtained $\hat{\Phi} = 0$ in this limit that is because of ignoring $\Delta_p$ and $\gamma_{bc}$. As we will obtain, these parameters have considerable effects on the propagation and storage of the probe pulse. If we substitute Eq. (21) in Eq. (14) we get a relation for atomic and field operators as a function of $\hat{\Psi}$ as below

$$\hat{\tilde{\sigma}}_{bc} = -\frac{1}{\sqrt{N}}(\sin\theta - \cos\theta$$

$$\times \frac{(i(\Delta+\Delta_p)+\gamma_{ba})(i\Delta_p+\gamma_{bc})\tan\theta\sin^2\theta}{g^2N+(i(\Delta+\Delta_p)+\gamma_{ba})(i\Delta_p+\gamma_{bc})\sin^2\theta})\hat{\Psi} , \quad (22a)$$

$$\hat{\mathcal{E}} = (\cos\theta + \sin\theta \times \frac{(i(\Delta+\Delta_p)+\gamma_{ba})(i\Delta_p+\gamma_{bc})\tan\theta\sin^2\theta}{g^2N+(i(\Delta+\Delta_p)+\gamma_{ba})(i\Delta_p+\gamma_{bc})\sin^2\theta})\hat{\Psi}.$$

(22b)

We see that if the behavior of $\hat{\Psi}$ is known, we can get easily the probe pulse behavior in any time. It can be inferred from Eq. (22b) that when we turn the control field off, the probe field does not tend to zero; instead, its amplitude may even be increase because of presence of $\tan\theta$ in Eq. (22b). This result shows that the corresponding result of Ref. [16] is not real (is an ideal result because of ignoring $\Delta_p$ and $\gamma_{bc}$.), where they have obtained zero value for the light field when control field is turned off.

**Proposition of reducing the control field intensity to small values instead of turning it to zero**

One may consider that when the control field intensity is absolutely turned to zero, $\tan\theta$ term and the expressions for $\hat{\Phi}, \hat{\mathcal{E}}$ (Eqs. (21,22b)) tend to infinity. The reason for this divergence in equations is that, when we turn the control field intensity to zero, the probe field increases and becomes comparable and even greater than the control field, causing the violation of the low intensity limit, which is assumed in deriving the above equations. Therefore, this divergence in equations occurs. The nonzero values for bright state ($\hat{\Phi}$) is not desirable, because it will cause an additional decay and loss of information (Ref.[2,16]), therefore, it is desired to avoid this phenomenon. It should be noted that this additional loss is not included and seen in our formulas, because we have used low intensity limit that is violated in this case. To avoid this phenomenon, we propose to reduce the control field intensity to small values instead of turning it to absolute zero. In this case $\tan\theta$ term will not have very large values and the expressions for $\hat{\Phi}, \hat{\mathcal{E}}$ will remain finite and small. Therefore, we will get the desired reduction of probe pulse velocity to have long storage times; also, we will avoid the creation of considerable values for bright state and the low intensity limit will not be violated. Therefore, since now, when we state turning off the control field, we mean to reduce the density of control field to small values such that the above conditions are complied.

Because all of coefficients in Eq. (22) are only functions of time, if we obtain the propagation of $\hat{\Psi}$, the same behavior will apply for the probe field and we can obtain it. Therefore, we define $\hat{\Psi}$ as the ***information pulse***, which can be totally the light field or totally the atomic excitation, according to the strength of control field. Information pulse contains the whole of information that is stored. We substitute Eq. (21) into Eq. (16) to reach a differential equation for only $\hat{\Psi}$ as below

-----

$$\frac{\partial}{\partial t}\hat{\Psi} + c(\cos^2\theta + B_0)\frac{\partial}{\partial z}\hat{\Psi} + [(i\Delta_p + \gamma_{bc})\sin^2\theta + A_0]\hat{\Psi} = 0 \qquad (23)$$

After some manipulation, we obtain relations for $A_0$ and $B_0$ as below

$$A_0 = \frac{(i(\Delta+\Delta_p)+\gamma_{ba})(i\Delta_p+\gamma_{bc})\tan\theta\sin^2\theta \times \dot{\theta} - (i(\Delta+\Delta_p)+\gamma_{ba})(i\Delta_p+\gamma_{bc})^2\sin^4\theta}{g^2N+(i(\Delta+\Delta_p)+\gamma_{ba})(i\Delta_p+\gamma_{bc})\sin^2\theta}, \qquad (24a)$$

$$B_0 = \frac{(i(\Delta + \Delta_p) + \gamma_{ba})(i\Delta_p + \gamma_{bc})\sin^4\theta}{g^2 N + (i(\Delta + \Delta_p) + \gamma_{ba})(i\Delta_p + \gamma_{bc})\sin^2\theta} \qquad (24b)$$

-------------------------------------------------------------------------------

As we will discuss later, $A_0$ term causes a loss and a phase shift of input pulse. $B_0$ term causes a modification of group velocity of information and light (probe) pulses. Therefore, when the control field is turned off (its intensity reduced to small values as stated in proposition of page 4), the group velocity of information pulse does not become zero. We have calculated this minimum velocity in the next section of the present paper. In addition, $B_0$ term causes a $k$-dependent loss (amplification) which results in dispersion and distortion of the light pulse.

It should be considered that our presented method is quantum mechanical and $\hat{\Psi}$ is an operator in the above equation; therefore, Eq. (23) governs the propagation of any quantum state of input probe pulse and can be used to study the storage of any quantum state in the memory. Therefore, the title "quantum memory" for this type of storage device is justified. To obtain the propagation of information pulse and therefore that of the light pulse in the medium, Eq. (23) that is a partial differential equation should be solved. To solve the present differential equation and to analyze further the propagation of light pulse, one may use Fourier Transformation (F.T.) method. It should be considered that $A_0, B_0$ and other coefficients in Eq. (23) are only functions of time, so that we can easily get Fourier transform of Eq. (23) with respect to z (space). In addition, one may consider that all coefficients are scalars and Fourier Transformation is just an integral transformation; therefore, we can extend the F.T. theory to the operators ($\hat{\Psi}$) and get the F.T. of Eq. (23) with considering $\hat{\Psi}$ as an operator. We consider the Fourier transform of $\hat{\Psi}$ as $\hat{\tilde{\Psi}}$ that is also an operator. We can also verify that the differentiation and shifting property of F.T. and all other operations that we will use later is valid for operators. Therefore, our treatment will be quantum mechanical in the following analysis. Only when we use the numerical calculation to study the propagation and storage of light pulse, our treatment becomes classical. Fourier Transformation is defined as

$$\hat{\tilde{\Psi}}(k,t) = \frac{1}{2\pi}\int_{-\infty}^{+\infty}\hat{\Psi}(z,t)e^{-ikz}dz. \qquad (25)$$

In which $\hat{\tilde{\Psi}}$ is the Fourier Transform of $\hat{\Psi}$ in momentum space. By applying F.T. to Eq. (23), we obtain an ordinary differential equation as

$$\frac{\partial}{\partial t}\hat{\tilde{\Psi}}(k,t) + [(i\Delta_p + \gamma_{bc})\sin^2\theta + A_0 + ikc(\cos^2\theta + B_0)]\hat{\tilde{\Psi}}(k,t) = 0. \qquad (26)$$

This equation can be solved by a simple integration and the solution is

$$\hat{\tilde{\Psi}}(k,t) = \hat{\tilde{\Psi}}(k,0)\exp[-\int_0^t [(i\Delta_p + \gamma_{bc})\sin^2\theta + A_0 + ikc(\cos^2\theta + B_0)]dt]. \qquad (27)$$

In which $\hat{\tilde{\Psi}}(k,0)$ is the F.T. of input information pulse (Polariton) at $t = 0$ (Corresponding to input light pulse with Eqs. (22)). Integrating the above equation
is difficult because $\theta$ is usually a complicated function of time that corresponds to the profile of turning on and off the control field. Therefore, we use the numerical methods to obtain output field from a given input field. In numerical calculation we switch to scalar (classical) values, $\mathcal{E}, \Psi$ and $\Phi$ that are expectation values of operators $\hat{\mathcal{E}}, \hat{\Psi}, \hat{\Phi}$ in the system. We calculate the scalar form of integral in Eq. (27), then get its Inverse Fourier Transform and then insert it in scalar counterpart of Eq. (22) to get $\mathcal{E}(z,t)$ in any time and location.

## 3. Results and Discussion

Before presenting the numerical results, we return to Eq. (27) to do some more analytical analysis on it. Eq. (27) can be rewritten in the following form

$$\hat{\tilde{\Psi}}(k,t) = \hat{\tilde{\Psi}}(k,0)\exp[-\int_0^t[\alpha_1 + i\beta + k\alpha_2 + ikv_g] \, dt \,] \qquad (28)$$

We can calculate $\alpha_1, \alpha_2, \beta, v_g$, by inserting expressions for $A_0, B_0$ into the Eq. (27). All of these coefficients are only functions of time. Eq. (28) is a valuable equation to understand the behavior of quantum memory and to predict the output light (probe pulse). We can now interpret every coefficient by considering Eq. (28). The $\alpha_1$ term determines the decay rate of the information pulse in every time and is the same for all $k$'s; therefore, it will not cause any dispersion or distortion of the information pulse and it will not cause the loss of information, but it only causes a decay of total pulse by the rate of $\alpha_1(t)$. The $\beta$ term corresponds to a phase shift of total information pulse. The $\alpha_2$ term is a $k$-dependent loss (Amplification) of information pulse that will cause to dispersion and distortion of information and light pulse, so that has the worst effect on storing information. $v_g$ is the velocity of information pulse in every time as we can infer it from the shifting property of F.T.

**Small detuning-High atomic density limit**

If we restrict ourselves to small detuning and high atomic densities, we can get our equations simpler. Therefore, we assume below condition

$$g^2 N \gg |(i(\Delta + \Delta_p) + \gamma_{ba})(i\Delta_p + \gamma_{bc})| \qquad (29)$$

By considering the typical values for parameters in the above equation, one finds that the present condition is usually fulfilled. In this case the denominator in $A_0$ and $B_0$ terms (Eq. (24)) reduces to $g^2 N$. We calculate the expressions for $\alpha_1, \alpha_2, \beta, v_g$ in this limit by reduced $A_0$ and $B_0$ terms and considering Eqs. (24,27,28) as below

$$\alpha_1 = \gamma_{bc}\sin^2\theta + \frac{\sin^2\theta}{g^2 N}[(\gamma_{bc}\gamma_{ba} - \Delta_p(\Delta+\Delta_p))(\tan\theta \times \dot\theta - \gamma_{bc}\sin^2\theta) + ((\Delta+\Delta_p)\gamma_{bc} + \Delta_p\gamma_{ba})(\Delta_p\sin^2\theta)] \qquad (30a)$$

$$\alpha_2 = -c((\Delta+\Delta_p)\gamma_{bc} + \Delta_p\gamma_{ba})\frac{\sin^4\theta}{g^2 N}, \qquad (30b)$$

$$\beta = \Delta_p \sin^2\theta + \frac{\sin^2\theta}{g^2 N}[((\Delta+\Delta_p)\gamma_{bc} + \Delta_p\gamma_{ba})(\tan\theta \times \dot\theta - \gamma_{bc}\sin^2\theta) - (\gamma_{bc}\gamma_{ba} - \Delta_p(\Delta+\Delta_p))(\Delta_p\sin^2\theta)], \qquad (30c)$$

$$v_g = c(\cos^2\theta + (\gamma_{bc}\gamma_{ba} - \Delta_p(\Delta+\Delta_p))\frac{\sin^4\theta}{g^2 N}). \qquad (30d)$$

The recent expressions are valuable relations as we can determine the properties of propagation and storage of information pulse in the medium by using them.

**a) Resonance condition**

We now simplify the above equations for the case of zero detuning ($\Delta = \Delta_p = 0$) and result are given as

$$\alpha_1 = \gamma_{bc}\sin^2\theta + \gamma_{bc}\gamma_{ba}(\tan\theta \times \dot\theta - \gamma_{bc}\sin^2\theta)\frac{\sin^2\theta}{g^2 N}, \qquad (31a)$$

$$\alpha_2 = 0, \qquad (31b)$$
$$\beta = 0, \qquad (31c)$$

$$v_g = c(\cos^2\theta + \gamma_{bc}\gamma_{ba}\frac{\sin^4\theta}{g^2 N}). \qquad (31d)$$

Usually the second term in $v_g$ is small compared to the first term, unless the control field is off (its density is reduced to small value compared to its initial value), such that the first term tends to zero. From the above equation, we can infer that when we turn off the control field ($\theta$ tends to $\pi/2$ and $\cos\theta \to 0$), we have the minimum velocity for information pulse (light pulse) in resonance case as

$$v_{g\min} = c\frac{\gamma_{bc}\gamma_{ba}}{g^2 N} \neq 0. \qquad (32)$$

It is considerable that we get a non-vanishing velocity for information pulse and it is because there is a non-vanishing $\gamma_{bc}$ in the system. Therefore, we infer that in quantum memory, which we are analyzing, storage does not mean in stoppage of light. The storage that takes place here, is only to reduce the speed of light that causes to trapping of light (information) for a long time in the memory cell. This procedure is of course different from trapping of light by setting the initial speed of light (when entering to the medium) to very small values (stationary slow light). The difference is that by turning off the control field and so that slowing down the speed of light when all of the probe pulse is inside the medium, we reduce the bandwidth of probe pulse. Therefore, its frequency components remain in the transparency window of EIT and therefore the pulse propagation remains adiabatic (the transparency window of EIT reduces by reducing the control field intensity.). The above equation can be regarded as a limit to the maximum storage time for a given length of the medium because the probe pulse will exit from the medium after a given time even when we have turned off the control field. The maximum storage time can be inferred from Eq.(32) and the length of the memory cell.

From Eqs.(31b),(31c), it can be inferred that there is no distortion and dispersion or even no phase shift for the information pulse when propagating inside medium in resonance case. Therefore, information stored in the memory remains undisturbed and we can get the same information in the output with only an attenuation in its amplitude. Therefore, we can increase the storage time to values as high as the limit that is inferred from Eq.(32) as stated above.

We see a very good agreement between the Eq. (32) and the numerical results that will be presented later. In high atomic density limit, we have $g^2 N \gg \gamma_{ba}\gamma_{bc}$ (Eq.(29)); therefore, we can neglect the third term in $\alpha_1$, so we get to an even simpler expression for $\alpha_1$ as

$$\alpha_1 = (\gamma_{bc} + \frac{\gamma_{bc}\gamma_{ba}}{g^2 N}\tan\theta \times \dot\theta)\sin^2\theta. \qquad (33)$$

**Slow light conditions**

If we set the initial speed of light pulse in the medium (when the control field is on) to be very smaller than the speed of light in vacuum, then $\theta$ will be very close to $\pi/2$ all the time and we can assume $\sin\theta$ equal to unity; therefore, Eq. (33) reduces to

$$\alpha_1 = \gamma_{bc} + \frac{\gamma_{bc}\gamma_{ba}}{g^2 N}\tan\theta \times \dot{\theta}. \tag{34}$$

The recent equation is the damping rate of information pulse for all times. We see that for $\dot{\theta}=0$ i.e. when the control field strength is constant, damping rate reduces to very simple relation

$$\alpha_1 = \gamma_{bc}. \tag{35}$$

That is equal to the damping rate of lower levels coherence. One may verify that the result of Eq. (35) is identical to the results of numerical calculations. (Fig. (3,4)). For $\dot{\theta}\neq 0$, $\dot{\theta}$ term portion to damping is small when control field has still a considerable value. i.e. when $\frac{\gamma_{ba}}{g^2 N}\tan\theta \times \dot{\theta} \ll 1$. ($\tan\theta$ is not very large). The $\dot{\theta}$ term will cause an additional decay rate only when $\Omega$ is changing and is very small. Therefore, its effect is considerable in very small times during the turn on and turn off of the control field and we can neglect its effect with some approximation on the overall damping of information pulse. Therefore, we can use an approximate, but very simple relation for the output probe pulse when it is coming out of the memory cell after a storage time of $T_0$ as below

$$\hat{\mathcal{E}}(t)|_{z_{out}} = \hat{\mathcal{E}}(t-T_0)|_{z_{in}} e^{-\gamma_{bc}T_0}. \tag{36}$$

Also $z_{out}$ and $T_0$ are related to each other by the equation below

$$z_{out} - z_{in} = \int_0^{T_0} v_g \, dt. \tag{37}$$

We see a good agreement between the results of numerical calculations (Fig. (3)) and the results obtained by Eq. (36) for damping of information pulse. Therefore, we can claim that approximation used for Eq. (36) is a proper approximation for usual cases. Of course, if we want to calculate the damping exactly, we may use the below equation to obtain the output field

$$\hat{\mathcal{E}}(t)|_{z_{out}} = \hat{\mathcal{E}}(t-T_0)|_{z_{in}} e^{-\int_0^{T_0}\alpha_1 dt}. \tag{38}$$

In which $\alpha_1$ is given by Eq. (34).

In addition, if we interested in calculating the output in a condition other than slow light condition i.e. for the case that our information pulse is propagating with a speed comparable to $c$ in the medium when the control field is on, we can use Eq. (33) for $\alpha_1$ in Eq. (38) to calculate the overall damping. With considering the above results, we see that if we ignore $\gamma_{bc}$ (that has the typical range of $10^2 - 10^5 \, rad/\sec$) in our relations (as it is done in previous work [16]), we reach to idealistic and inexact results about propagation and storage of light in the memory.

**b) Off- resonance condition**

In the off-resonance case, $\alpha_2,\beta$ becomes nonzero in general. $\beta$ corresponds to a phase shift of total information pulse and do not considerably affect on destroying of stored information. However, nonzero $\alpha_2$, if it is considerable, will cause a $k$-dependent loss (amplification) leading to distortion and dispersion (fast oscillations) of the light pulse (as it is also seen in numerical calculations). In quantum memory, we should minimize any distortion and dispersion because it destroys the stored information; therefore, we should try to reduce the value of $\alpha_2$. We can see from Eq. (30b), that if $\Delta, \Delta_p$ have different signs, they will reduce effects of each other and even they can be adjusted to yield $\alpha_2 = 0$. However, this adjustment will cause an additional decay rate and increased minimum group velocity that is not desired. In addition, we can set $v_{g\min}$ in Eq. (30d) to be zero by setting $\Delta, \Delta_p$ to proper values, however it will cause a considerable value of $\alpha_2$ (regarding Eq. (30b)) that destroys completely the stored information. We also try to decrease $\alpha_1$ (the decay rate) compared to resonance case, by setting $\Delta, \Delta_p$ in Eq. (30a). However, we see that every attempt to reduce $\alpha_1$ causes in considerable value of $\alpha_2$ and destroys the stored information. Therefore, we deduce that any detuning (in small detuning and high atomic density limit, Eq. (29)) is undesired in quantum memory and will result in distortion and loss of stored information. Therefore, we should try to set the system in resonance. In addition, we deduce that the system is much more sensitive to $\Delta_p$ than $\Delta$ (consider Eq. (30b)) and a small $\Delta_p$ (comparable to $0.01\frac{g^2 NL_p}{c\gamma_{ba}T_0}$) will cause a complete loss of information. ($T_0$ is the storage time and $L_p$ is the length of pulse in the medium.)

We derive limitations to $\Delta, \Delta_p$ in order to have acceptable (undistorted) output from Eq. (28,30b) as follows

$$\Delta_p < 0.01\frac{g^2 NL_p}{c\pi\gamma_{ba}T_0}, \tag{39a}$$

$$\Delta < 0.01\frac{g^2 NL_p}{c\pi\gamma_{bc}T_0}. \tag{39b}$$

It should be considered that in our model, we have assumed a single carrier frequency for the probe and coupling fields that are idealistic and impractical. [This assumption is common (as it is assumed in previous works), because it causes an enormous simplification in analysis.] However, in practical situation, our lasers have a considerable bandwidth and we cannot reduce their bandwidths more than a specified range. We may consider the system of laser fields with finite bandwidths as a set of probe and coupling fields with different carrier frequencies applied to atomic system. This causes to presence and action of various $\Delta, \Delta_p$'s on the system; therefore, there will be an undesired distortion and loss of stored information that cannot be eliminated. We deduce from Eq. (39) and the above discussion that our lasers should have narrow and very close to each other bandwidths, also, they should have center frequencies close to resonances, to have an acceptable efficiency for quantum storage of light. These conditions are given by

$$|BW_c - BW_p|, |(\omega_{ab} - \nu_{0p}) - (\omega_{ac} - \nu_{0c})| < 0.01 \frac{g^2 N L_p}{c \pi \gamma_{ba} T_0} \quad (40a)$$

$$BW_{c,p} < 0.01 \frac{g^2 N L_p}{c \pi \gamma_{bc} T_0}. \quad (40b)$$

In the recent conditions, $BW_\mu, \nu_{0\mu}$ are the bandwidth and the center frequencies of the corresponding lasers. Eq.(40) may also be considered as a limit to maximum storage time of information in the memory. We can infer from Eq.(40) that for a given set of practical lasers with finite bandwidths we can not increase the storage time upper than a maximum value with undisturbed output. Therefore, we have set two limits to the maximum storage time of practical quantum memory as given by Eq.(32) and Eq.(40).

### 4. Numerical Calculations

We have calculated the scalar counterpart of integral in Eq. (27) and other procedure numerically to obtain the Information Pulse and light (probe) pulse in any time and space. We use the typically reported values for the parameters in this numerical calculation [7-9,24-26].

We set the parameters as follow. $L = 5^{mm}$ (Length of the memory cell); $D = 200^{\mu m}$ (Diameter of the Cell), (the volume of the cell is derived from these two parameters); $\wp_{ca} \approx \wp_{ba} \approx 10^{-29} - 10^{-28}$ C.m (Electric dipole of corresponding levels); $\nu_p = 5 \times 10^{14} Hz \times 2\pi$ (Probe field frequency); $g = 10^6 \, rad/\sec$ (Vacuum Rabi frequency); $N = 10^8 \, atoms$ (Number of atoms in the cell corresponding to the atomic density of $10^{12} cm^{-3}$). In the following calculations, we use the input information pulse as

$$\Psi(z, t=0) = 0.2 \times e^{-(\frac{z}{10^{-3}})^2} \quad (41)$$

That corresponds to the probe field intensity of $I_p \cong 5.3 \times 10^{-3} W/m^2$. In addition, we use a profile to turn off and turn on the control field as follows:

$$\theta = Arc \cot[5 \times 10^{-4} \times \{1 - 0.5 \times \tanh(10^5 \times (t - t_1)) + 0.5 \times \tanh(10^5 \times (t - t_2))\} + 10^{-5}] \quad (42)$$

The parameters $t_1, t_2$ are the turn off and turn on times that we set to be $t_1 = 30 \times 10^{-6}$ sec, $t_2 = 125 \times 10^{-6}$ sec. Indeed, this profile does not turn absolutely the control field intensity to zero, instead it reduces the control field intensity to a small value ($\Omega = 10^5 \, rad/\sec$). When we state the turning off of the control field, we mean to reduce its intensity to a very small value compared to its initial value (on time intensity) as stated in proposition of page 4. Therefore, the divergence of Eqs. (21,22b) will not occur and the bright state and the corresponding decay and loss will always have negligible values; also, the low intensity limit will not be violated and our formulas will still be valid. This small value for control field intensity in the off times, will cause a small increase in the group velocity of the probe pulse (0.03 m/sec), that is negligible compared to $v_{gmin}$.

We have shown the $\theta, \Omega$ in the Fig. (2) as a function of time as given by above equation. We have set by the above equation the Rabi frequency of control field (Control field strength) in the "on time" to be about $\Omega_c = 5 \times 10^6$ rad/sec (Its intensity is set to about $I_c \cong 132.5 \times 10^{-3} W/m^2$). It results in the initial velocity of light in the medium equal to about (75+ $v_{gmin}$) m/sec (Slow initial velocity). As one may infer $\theta$ is very close to $\pi/2$ even when the control field is on. In this case, the dominant part of information pulse is the atomic excitations in all times. Setting the initial velocity of light pulse to be small is reasonable (from practical viewpoint) to trap the light (as it is done in experimental works [24,25]). If we set the initial velocity of light near speed of light in vacuum, it would escape the medium before the control field is turned off and its control would be difficult. We have also plotted the velocity of Information pulse for resonance case in Fig.(2,c,d) for different parameters using Eq. (31d). One should remember that all of the following simulation is done by using the parameters and $\theta(t)$ and $\Psi(z, t=0)$ profiles given above.

**a) Resonance case**

Fig.(3,a-d) shows the numerical results of information pulse propagation and storage for time steps of $15 \mu$ sec and given values of parameters. Fig.(3,e,f) shows the peak value of information pulse versus time for given values of parameters. One may verify that the results in Eqs. (35,36) for damping of information pulse are identical to the numerical results. Fig. (4) shows the information pulse in storage region where the control field is off (storage region in Fig (3)). One may easily verify the minimum group velocity from Eq. (32) (that yields a: $v_{gmin} = 3 m/\sec$, b: $v_{gmin} = 0.3 m/\sec$, c: $v_{gmin} = 30 m/\sec$, d: $v_{gmin} = 3 m/\sec$) and the

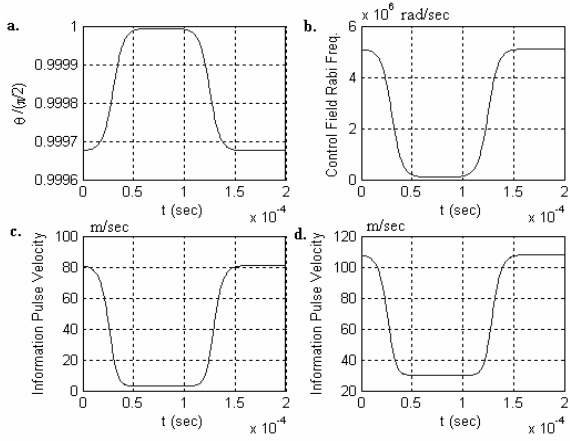

**Fig. 2.** a) $\theta/\frac{\pi}{2}$ as a function of time. b) Rabi frequency of control field as a function of time. c) Group velocity of information pulse for $\gamma_{ba}=10^8$, $\gamma_{bc}=10^4$ ($rad/\sec$), $\Delta,\Delta_p=0$. d) Group velocity of information pulse for $\gamma_{ba}=10^9$, $\gamma_{bc}=10^4$ ($rad/\sec$), $\Delta,\Delta_p=0$. (c and d are plotted using Eq. (31d)).

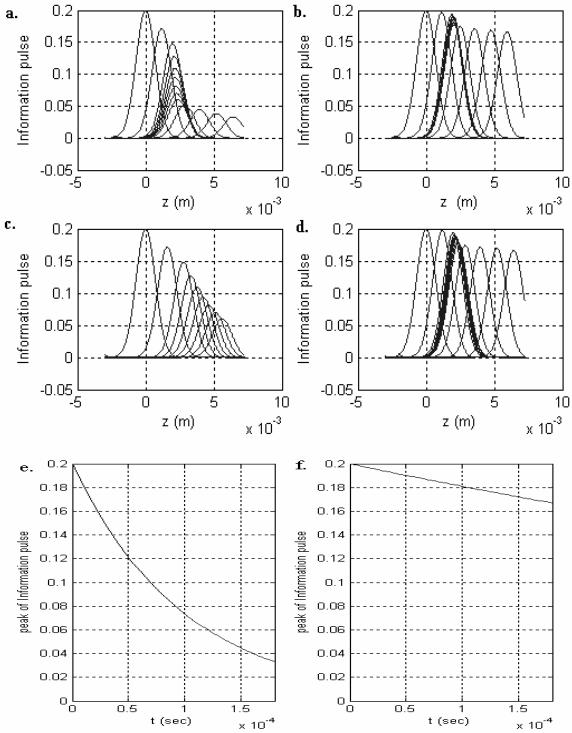

**Fig.3.** Information pulse propagation for time steps of $15\,\mu\sec$ from $t=0$ to $t=180\,\mu\sec$, with $\theta(t)$ (turn on and off) profile given by Eq.(42) .a) $\gamma_{ba}=10^8$, $\gamma_{bc}=10^4$ ($rad/\sec$). b) $\gamma_{ba}=10^8$, $\gamma_{bc}=10^3$ ($rad/\sec$). c) $\gamma_{ba}=10^9$, $\gamma_{bc}=10^4$ ($rad/\sec$) (From $t=0$ to $t=120\mu\sec$) d) $\gamma_{ba}=10^9$, $\gamma_{bc}=10^3$ ($rad/\sec$). e) Peak (maximum) of information pulse versus time for $\gamma_{ba}=10^8$, $\gamma_{bc}=10^4$ ($rad/\sec$). f) Peak (maximum) of information pulse versus time for $\gamma_{ba}=10^8$, $\gamma_{bc}=10^3$ ($rad/\sec$). ($\Delta,\Delta_p=0$).

damping rate from Eq. (35) are identical to numerical results. (Remember that $v_{gmin}$ depends on $\gamma_{bc}$ and $\gamma_{ba}$, however damping rate depends only on $\gamma_{bc}$ with respect to Eqs. (32,35)). We see that reducing the control field intensity to small values instead of turning it to absolute zero according to Eq.(42), have no considerable effect on the group velocity of information pulse, compared to the absolute turn off. As it was stated, it increases the group velocity by 0.03 m/sec that is negligible compared to $v_{gmin}$. We can see the differences between the results of the previous works [16] and our results. In the previous works, there is no decay for the information pulse (Polariton) in adiabatic passage limit and the information pulse is slowed down to stop when control field is off.

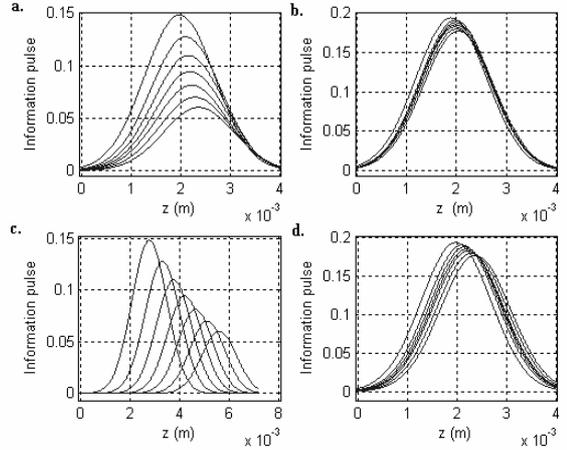

**Fig. 4.** Information pulse for time steps of $15\,\mu\sec$ in the storage region where the control field is off (from $t=30\,\mu\sec$ to $t=120\mu\sec$) with $\theta(t)$ (turn on and off) profile given by Eq.(42) a) $\gamma_{ba}=10^8$, $\gamma_{bc}=10^4$ ($rad/\sec$). b) $\gamma_{ba}=10^8$, $\gamma_{bc}=10^3$ ($rad/\sec$). c) $\gamma_{ba}=10^9$, $\gamma_{bc}=10^4$ ($rad/\sec$). d) $\gamma_{ba}=10^9$, $\gamma_{bc}=10^3$ ($rad/\sec$). ($\Delta,\Delta_p=0$).

Fig.5 shows the bright state pulse for the time steps of 15 $\mu\sec$. As it is shown, the bright state grows to high values when the control field is off. It is because of presence of $\tan\theta$ in the Eq.(21). Because the bright state in off time of control field is very higher than that in other times, in Fig.(b),(d) bright state for other times are very small that can not be seen. We show initial bright state at $t=0$ and output bright state at $t=165\mu\sec$ for $\gamma_{ba}=10^8$, $\gamma_{bc}=10^4$ $rad/\sec$, in Fig.(5-a,c) to compare it with bright state at other times. (Pay attention to the scaling in figures). We see that reducing the control field intensity to small values instead of turning it to absolute zero, have caused the bright state to remain finite and very smaller than the dark state ($\Psi$) in off times; therefore, we deduce that our proposition is efficient to eliminate the divergence of equations and additional losses. It may be considered that in previous works [16], the bright state is always zero valued in the adiabatic passage limit and their formulation do not show any increase in bright state, in off times of control field that is because of ignoring $\gamma_{bc}$ in their formulation.

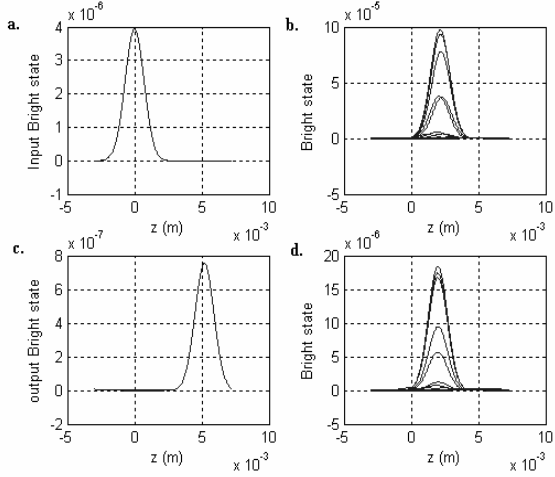

**Fig.5.** a) Initial bright state for input pulse and $\gamma_{ba} = 10^8$, $\gamma_{bc} = 10^4$ ($rad/\sec$). b) Bright state for time steps of $15\mu\sec$ for $\gamma_{ba} = 10^8$, $\gamma_{bc} = 10^4$ ($rad/\sec$). c) Output Bright state at time $t = 165\mu\sec$ and for $\gamma_{ba} = 10^8$, $\gamma_{bc} = 10^4$ ($rad/\sec$). d) Bright state for time steps of $15\mu\sec$ and for $\gamma_{ba} = 10^8$, $\gamma_{bc} = 10^3$ ($rad/\sec$). ($\Delta, \Delta_p = 0$ and with $\theta(t)$ (turn on and off) profile given by Eq.(42)).

Fig.6 shows the propagation of light (probe) pulse ($\mathcal{E}(z)$) for the time steps of $15\mu\sec$. We see that the velocity for the probe pulse is identical to the velocity of information pulse and the information is turned back to the probe field as it was at the initial time ($t = 0$), when we turn on the control field again. We see that when the control field is turned off (with the profile of $\theta(t)$ given by Eq.(42)), the probe field does not tend to zero. As we see from the figure, the peak of the light pulse remains finite in the range of $\mathcal{E} \approx 10^{-4}$ in off times of the control field. This value corresponds to the Rabi frequency of light field about $\Omega_p \approx 10^2 \, rad/\sec$ that is very smaller than $\Omega_c$ in off times ($10^5$); therefore, we see that the low intensity limit remains valid in this case. This is because of using the profile of $\theta(t)$ as given by Eq. (42), that reduces the control field intensity to small value instead of turning it to absolute value. There is an appear difference between the present results and the results of Ref.[16] in which the probe field tends to zero and is completely converted to atomic excitation when the control field is turned off. This difference is because of ignoring $\gamma_{bc}$ in their formulation that has made their approach ideal and inexact.

Fig.7 shows the atomic excitation ($\sigma_{bc}$) for various times. We see that it only decays by the rate of $\alpha_1$ and when we turn off the control field, there is no considerable variation (In contrast with results of [16]). This difference is not fundamental and is because we have set the initial velocity of the light pulse in a very small value (as it is done in practical reports [24,25]) and dominant part of the information pulse in the medium is atomic excitation since it enters the medium.

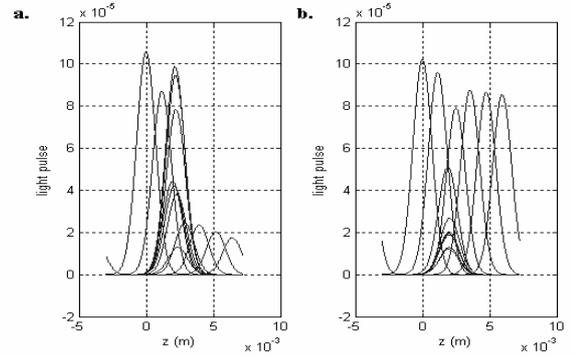

**Fig.6** a) light pulse ($\mathcal{E}(z)$) for time steps of $15\mu\sec$ for $\gamma_{ba} = 10^8$, $\gamma_{bc} = 10^4$ ($rad/\sec$). b) light pulse for time steps of $15\mu\sec$ and for $\gamma_{ba} = 10^8$, $\gamma_{bc} = 10^3$ ($rad/\sec$). ($\Delta, \Delta_p = 0$ and with $\theta(t)$ (turn on and off) profile given by Eq.(42)).

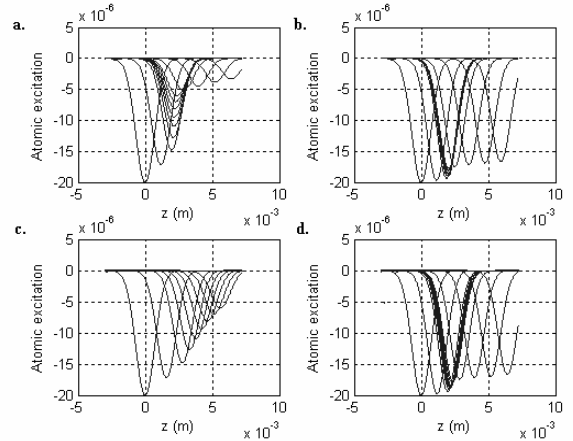

**Fig.7.** $\sigma_{bc}$ (atomic excitation) propagation for time steps of $15\mu\sec$ from $t = 0$ to $t = 180\mu\sec$ with $\theta(t)$ (turn on and off) profile given by Eq.(42). a) $\gamma_{ba} = 10^8$, $\gamma_{bc} = 10^4$ ($rad/\sec$). b) $\gamma_{ba} = 10^8$, $\gamma_{bc} = 10^3$ ($rad/\sec$). c) $\gamma_{ba} = 10^9$, $\gamma_{bc} = 10^4$ ($rad/\sec$) (From $t = 0$ to $t = 120\mu\sec$). d) $\gamma_{ba} = 10^9$, $\gamma_{bc} = 10^3$ ($rad/\sec$). ($\Delta, \Delta_p = 0$).

## b) Off-resonance case

Fig. (8,9) shows the input information pulse and output information pulse at the time $t = 165\mu\sec$ for various $\Delta, \Delta_p$'s. We can now examine the limitation in Eq. (39) to be correct. If we substitute the parameters of simulation for Fig. (8,9) into the Eq. (39), we reach to the limits of $\Delta_p = 2\times 10^2 \, rad/\sec$ and $\Delta = 2\times 10^6 \, rad/\sec$ for detunings. We see that when the condition in Eq. (39a) or Eq. (39b) is violated, the output information pulse bears fast oscillations and gets destroyed that causes the loss of total information (Part c and e in fig. (8,9)). As stated before, this is because we have a *k*-dependent loss (amplification) in the Eq. (28). With the values used in calculations, we see that any attempt to reduce $v_{gmin.}, \alpha_1$ in Eq. (30) with setting $\Delta, \Delta_p$, will cause a complete loss of information. In the recent figures, we show both the real and imaginary parts of the information pulse. We see that in the off resonance case, the imaginary part of the output information pulse becomes nonzero that is caused by nonzero $\beta$. This effect causes a phase shift in the output light pulse. It should be noted that this phase shift depends on the values of detunings and the storage time.

From Fig. (8,9), it can be inferred that if our practical lasers are not well adjusted and do not have very narrow bandwidths, we will not be able to have any storage of information. We find that Eq.(40) sets a very small and strict limit for the Bandwidth and center frequency of lasers and these values are difficult to achieve by conventional lasers and optical devices, therefore for practical conditions the maximum storage time is very small.

We here refer to Ref. [17]. In that paper, they have set a limit to Two-photon detuning $\Delta_p$ (that is referred as $\delta$ there (Eq. (23)). Their limitation is far larger than the limit that we have obtained in Eq. (39) and is not the effective limit because they do not consider $\gamma_{bc}$; also, they apply the two photon detuning as a perturbation and derive the limitation in the first order of this perturbation.

## V) Conclusion

In this paper we further developed the quantum mechanical theory for quantum light memory in the *Low intensity limit* and *adiabatic passage limit*, primarily developed by Ref.[16]. We entered the parameters $\gamma_{bc}, \Delta, \Delta_p$ into the formulations. We obtained and explained their effects in a clear form. We analyzed the propagation and storage of the light pulse in the resonance case and obtained the decay rate and the minimum group velocity in this case. In addition, we reached to a non-vanishing light field when the control field is turned off in storage process and we proposed to reduce the control field intensity instead of turning it to absolute zero to avoid additional losses caused by large values for bright state. We then analyzed the off-resonance case and reached to the result that off-resonance case (In Small detuning and High atomic density limit) has no advantage and can destroy completely the output (stored) information. We obtained limitations for maximum value of detunings and therefore, limitations for bandwidths and center frequencies of practical lasers used, in order to maintain the stored information from distortion. In addition, we set limitations for maximum storage time of practical quantum memory. We then presented the related numerical results to verify the analytical results and limits.

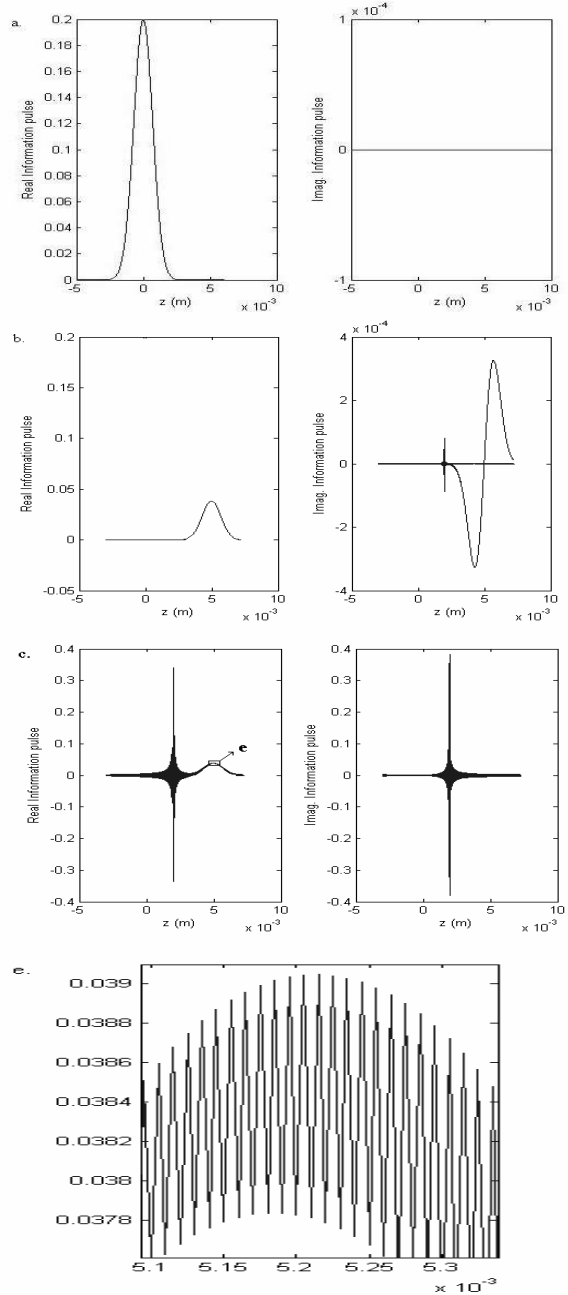

**Fig.8.** Real and Imaginary parts of the information pulse a) Input information pulse (at the time $t = 0$). b) Output information pulse (at $t = 165\mu\sec$) for $\Delta = 2 \times 10^6$ ($rad/\sec$). c) Output information pulse (at $t = 165\mu\sec$) for $\Delta = 5 \times 10^6$ ($rad/\sec$). e) Enlarged peak of real part of information pulse in part c of figure (shows the fast oscillations in output). ($\Delta_p = 0, \gamma_{ba} = 10^8, \gamma_{bc} = 10^4 (rad/\sec)$, with $\theta(t)$ (turn on and off) profile given by Eq.(42)).

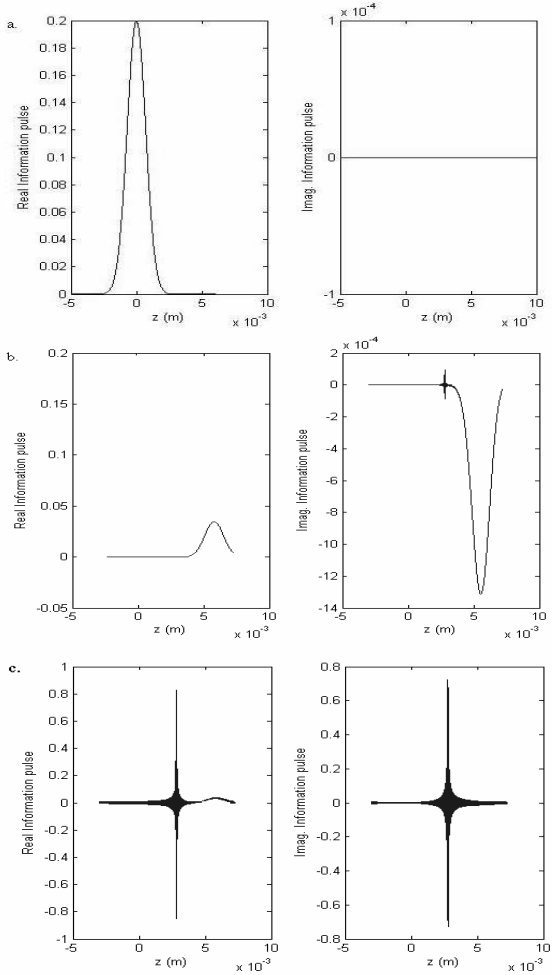

**Fig.9.** Real and Imaginary parts of the information pulse a) Input information pulse (at the time $t = 0$ ). b) Output information pulse (at $t = 165\mu\sec$ ) for $\Delta_p = 2 \times 10^2$ ( $rad/\sec$ ). c) Output information pulse (at $t = 165\mu\sec$ ) for $\Delta_p = 5 \times 10^2$ ( $rad/\sec$ ). ( $\Delta = 0, \gamma_{ba} = 10^8, \gamma_{bc} = 10^4$ ($rad/\sec$) , with $\theta(t)$ (turn on and off) profile given by Eq.(42)).